\documentclass[12pt]{article}   

\usepackage{amsbsy}
\usepackage{amsmath}
\usepackage{amssymb}
\usepackage{amsfonts}
\usepackage{graphics}
\usepackage{epsfig}

\newlength{\dinwidth}                       
\newlength{\dinmargin}                      
\newcommand{\gap}{\stackrel{>}{\sim}}

\newcommand{\half}{\mbox{\small $\frac{1}{2}$}}
\newcommand{\quarter}{\mbox{\small $\frac{1}{4}$}}

\newcommand{\Dd}[1]{\mbox{
  \parbox[b]{0cm}{$D$}\raisebox{1.7ex}{$\leftrightarrow$}$_{\!#1}$}}

\def\lsim{\mathrel{\rlap{\lower4pt\hbox{\hskip1pt$\sim$}}
    \raise1pt\hbox{$<$}}}                
\def\gsim{\mathrel{\rlap{\lower4pt\hbox{\hskip1pt$\sim$}}
    \raise1pt\hbox{$>$}}}                
\begin{document}

\title{
\vspace{-2.5cm}
\flushleft{\normalsize DESY 99-129} \hfill\\
\vspace{-0.35cm}
{\normalsize HLRZ 99-37} \hfill\\
\vspace{-0.35cm}
{\normalsize HUB-EP-99/40} \hfill\\
\vspace{-0.35cm}
{\normalsize TPR 99-16} \hfill\\
\vspace{-0.35cm}
{\normalsize August 1999} \hfill\\
\vspace{0.5cm}
\centering{\LARGE \bf
The Polarized Structure Function 
$\boldsymbol{g_2}$:\\ A Lattice Study 
Revisited\footnote{Talk given by G. Schierholz at `Workshop on {\it Polarized
    Protons at High Energies -- Accelerator Challenges and Physics
    Opportunities}', 17 - 20 May 1999, DESY, Hamburg, Germany.}}
\\[1em]}
\author{\large M. G\"ockeler$^1$, R. Horsley$^2$, W. K\"urzinger$^{3,4}$,\\
H. Oelrich$^3$, P. Rakow$^1$ and G. Schierholz$^{3,5}$\\[2em]
        $^1$ Institut f\"ur Theoretische Physik, Universit\"at Regensburg,\\
                    D-93040 Regensburg,
                    Germany\\[0.5em]
        $^2$ Institut f\"ur Physik, Humboldt-Universit\"at zu Berlin,\\
                    D-10115 Berlin, Germany\\[0.5em]
        $^3$ Deutsches Elektronen-Synchrotron DESY,\\ 
             John von Neumann-Institut f\"ur Computing NIC,\\ 
                    D-15735 Zeuthen, Germany\\[0.5em]
        $^4$ Institut f\"ur Theoretische Physik, Freie Universit\"at Berlin,\\
                    D-14195 Berlin, Germany\\[0.5em]
        $^5$ Deutsches Elektronen-Synchrotron DESY,\\ 
                    D-22603 Hamburg, Germany}

\date{ }

\maketitle

\begin{abstract}
A recent lattice calculation of the spin-dependent structure function $g_2$ 
is revisited. It has been recognized that the twist-three operator, which
gives rise to $d_2$, mixes non-perturbatively with operators of lower
dimensions under renormalization. This changes the results substantially.
\vspace*{0.7cm}
\end{abstract}

\section{Introduction}

The nucleon's second spin-dependent structure function $g_2$ is of 
considerable phenomenological interest. The discussion follows the
operator product expansion (OPE)~\cite{Jaffe}. In leading order 
of $1/Q^2$, $g_2$ receives contributions from both, twist-two and twist-three
operators. It thus offers the unique possibility of directly assessing
higher-twist effects. The twist-three operator probes the
transverse momentum distribution of the quarks in the nucleon, and has
no simple parton model interpretation.

A few years ago we have computed the lowest non-trivial moment of $g_2$ on
the lattice~\cite{QCDSF1}. To convert the lattice numbers to continuum
results, the operators entering the OPE have to be renormalized. An essential 
feature of our calculation was that the renormalization was done in 
perturbation theory. It turned out that the twist-three contribution was the
dominant contribution to both, the proton and the neutron structure functions.

Renormalization effects are a major source of systematic error. In this talk 
we shall extend our previous work by employing non-perturbative 
renormalization. The novel feature of this approach is that it introduces
mixing with lower-dimensional operators.

\section{OPE and Mixing}

In leading order of $1/Q^2$, and for massless quarks, the moments of $g_2$ 
are given by
\begin{equation}
\begin{split}
2\int_0^1\mbox{d}x x^n g_2(x,Q^2)  
  = \frac{1}{2}\frac{n}{n+1} \sum_{f=u,d} &\big[e^{(f)}_{2,n}(\mu^2/Q^2,g(\mu))
\: d_n^{(f)}(\mu) \\
&-e^{(f)}_{1,n}(\mu^2/Q^2,g(\mu))\: a_n^{(f)}(\mu)\big]
\end{split}
\label{mom}
\end{equation}
for even $n \ge 2$, 
where \cite{Jaffe}
\begin{eqnarray}
\langle \vec{p},\vec{s}| 
 {\cal O}^{5 (f)}_{ \{ \sigma\mu_1\cdots\mu_n \} }
                       | \vec{p},\vec{s} \rangle 
   &=& \frac{1}{n+1}a_n^{(f)} \,[ s_\sigma p_{\mu_1} \cdots p_{\mu_n} 
+ \cdots -\mbox{traces}], \label{twist2} \\
\langle \vec{p},\vec{s}| 
 {\cal O}^{5 (f)}_{ [  \sigma \{ \mu_1 ] \cdots \mu_n \} }
                       | \vec{p},\vec{s} \rangle 
   &=& \frac{1}{n+1}d_n^{(f)} \,[ (s_\sigma p_{\mu_1} - s_{\mu_1} p_\sigma)
 p_{\mu_2}\cdots p_{\mu_n} + \cdots -\mbox{traces}], 
\label{twist3} \\
 {\cal O}^{5 (f)}_{\sigma\mu_1\cdots\mu_n}
   & =& \left(\frac{i}{2}\right)^n\bar{\psi}\gamma_{\sigma} \gamma_5
 \mbox{\parbox[b]{0cm}{$D$}\raisebox{1.7ex}{$\leftrightarrow$}}_{\mu_1}  
 \cdots \mbox{\parbox[b]{0cm}{$D$}\raisebox{1.7ex}{$\leftrightarrow$}}_{\mu_n}
 \psi -\mbox{traces},
\end{eqnarray}
and $e_{1,n}^{(f)}$, $e_{2,n}^{(f)}$ are the Wilson coefficients.
Here $\{\cdots\}$ ($[\cdots]$) means symmetrization
(antisymmetrization). The operator (\ref{twist2}) has twist two, whereas
the operator (\ref{twist3}) has twist three.
Both, the Wilson coefficients and the operators are renormalized at the
scale $\mu$. It is assumed that the Wilson coefficients can be computed 
perturbatively, whereas the calculation of 
the reduced matrix elements $a_n^{(f)}$ and $d_n^{(f)}$ is a problem for the 
lattice. In the following we shall drop the flavor indices, unless they are 
necessary.

The lattice calculation splits into two separate tasks. The first task is to
compute the nucleon matrix elements of the appropriate lattice operators. This
was described in detail in~\cite{QCDSF1}. The second task
is to renormalize the operators. Generically,  
\begin{equation}
   {\cal O}(\mu) = Z_{\cal O}((a\mu)^2, g(a))\, {\cal O}(a).
\label{op1}
\end{equation}
As in the continuum, we impose the (MOM-like) renormalization condition
\begin{equation}
\quarter \,\mbox{Tr} \, \langle q(p)|{\cal O}(\mu)|q(p)\rangle 
\Big[\langle q(p)|{\cal O}(a)|q(p)\rangle\, |^{\rm tree}\Big]^{-1}
\underset{p^2 =\mu^2}{=} 1,
\end{equation}
where $|q(p)\rangle$ is a quark state of momentum $p$ in Landau gauge.
In our earlier work~\cite{QCDSF1,QCDSF2} we have computed the renormalization
constants in perturbation theory to one-loop order. A restriction of the
perturbative calculation is that it does not allow to incorporate mixing with
lower-dimensional operators.

In a recent paper~\cite{QCDSF3} we have started a non-perturbative 
calculation of the renormalization constants associated with the 
structure functions $F_1$, $F_2$ and $g_1$.  
Let us here consider the case of the structure function $g_2$.
We shall restrict ourselves to $n = 2$. This is the 
lowest moment of $g_2$ for which the OPE makes a statement.  
As before~\cite{QCDSF1}, we take $\sigma = 2$, $\mu_1=1$ and $\mu_2=4$. 
We thus need to consider the operators
\begin{equation}
\bar{\cal O}^5_{\{214\}} =: {\cal O}^{\{5\}}
\label{os}
\end{equation}
and
\begin{eqnarray}
\bar{\cal O}^5_{[2\{1] 4\}} &=&  2 \bar{\cal O}^5_{2\{14\}} 
- \bar{\cal O}^5_{1\{24\}} -  \bar{\cal O}^5_{4\{12\}} \nonumber \\   
&=&\bar{\psi}\Big(\gamma_2 \Dd{1} \Dd{4} + \gamma_2 \Dd{4} \Dd{1} 
   - \half \gamma_1 \Dd{2} \Dd{4} - \half \gamma_1 \Dd{4} \Dd{2} \nonumber \\
& & \quad \quad \quad - \half \gamma_4 \Dd{1} \Dd{2} - \half \gamma_4 \Dd{2}
\Dd{1}\Big) \gamma_5 \psi \nonumber \\
&=:& {\cal O}^{[5]}, \label{o5}
\end{eqnarray}
with $\bar{\cal O}^5$ indicating the Euclidean counterpart of ${\cal O}^5$,
which belong to the representation $\tau_3^{(4)}$ and 
$\tau_1^{(8)}$, respectively, of the hypercubic group
$H(4)$~\cite{Mandula}. The operator (\ref{o5}) has dimension five and
$C$-parity $+$. It turns out that there exist two operators of dimension
four and five, respectively, transforming identically 
under $H(4)$ and having the same $C$-parity, with which (\ref{o5}) can mix:
\begin{equation}
\mbox{i}\, 
 \bar{\psi} \Big(\sigma_{13} \Dd{1} -  \sigma_{43} \Dd{4}\Big) \psi
=: {\cal O}^\sigma, \label{osigma}
\end{equation}
\vspace*{-0.7cm}
\begin{equation}
\bar{\psi} \Big(\gamma_1 \Dd{3} \Dd{1} - \gamma_1 \Dd{1}
 \Dd{3} - \gamma_4 \Dd{3} \Dd{4} + \gamma_4 \Dd{4} \Dd{3}\Big) \psi =:
{\cal O}^0. \label{o0}
\end{equation}
The operator (\ref{o0}) vanishes in tree approximation between quark states. 
It therefore cannot be included in our framework, and so we discard it.
We then remain with
\begin{equation}
{\cal O}^{[5]}(\mu) = Z^{[5]}(a\mu) {\cal O}^{[5]}(a) + Z^\sigma(a\mu) 
{\cal O}^\sigma(a).
\label{renorm}
\end{equation}
The renormalization constant $Z^{[5]}$ and the mixing coefficient $Z^\sigma$
are determined from
\begin{equation}
\quarter \,\mbox{Tr} \,\langle q(p)|{\cal O}^{[5]}(\mu)|q(p)\rangle 
\Big[\langle   q(p)|{\cal O}^{[5]}(a)|q(p)\rangle\, |^{\rm tree}\Big]^{-1} 
\underset{p^2 =\mu^2}{=}
  1, \label{cond1}
\end{equation}
\vspace*{-0.7cm}
\begin{equation} 
\quarter \,\mbox{Tr} \,\langle q(p)|{\cal O}^{[5]}(\mu)|q(p)\rangle 
\Big[\langle
  q(p)|{\cal O}^\sigma(a)|q(p)\rangle\, |^{\rm tree}\Big]^{-1}
\underset{p^2 =\mu^2}{=} 0. \label{cond2} 
\end{equation}
Let us now see how much this effect changes our results for $d_2$. 
 
\section{Results and Conclusions}

We work on a $16^3\, 32$ lattice at $\beta = 6.0$ using quenched Wilson
fermions. The lattice spacing is $a^{-1} \approx 1.95 \, \mbox{GeV}$. The 
calculations are done at three values of the hopping parameter $\kappa$, so 
that we can extrapolate our results to the chiral limit. We denote the reduced
matrix elements of the unrenormalized operators ${\cal O}^{[5]}$ and
${\cal O}^\sigma$
by $d_2^{[5]}$ and $d_2^\sigma$, respectively. In table \ref{me} we give our 
results. The numbers
for $a_2$ and $d_2^{[5]}$ (formerly called $d_2$) are taken over
from~\cite{QCDSF1}. These numbers are based on $O(400-1000)$ configurations,
depending on the value of $\kappa$. The results for $d_2^\sigma$ are new, and
they are based on $O(80)$ configurations only. Hence the relatively large
statistical error.

\begin{figure}[htbp]
  \begin{center}
    \epsfig{file=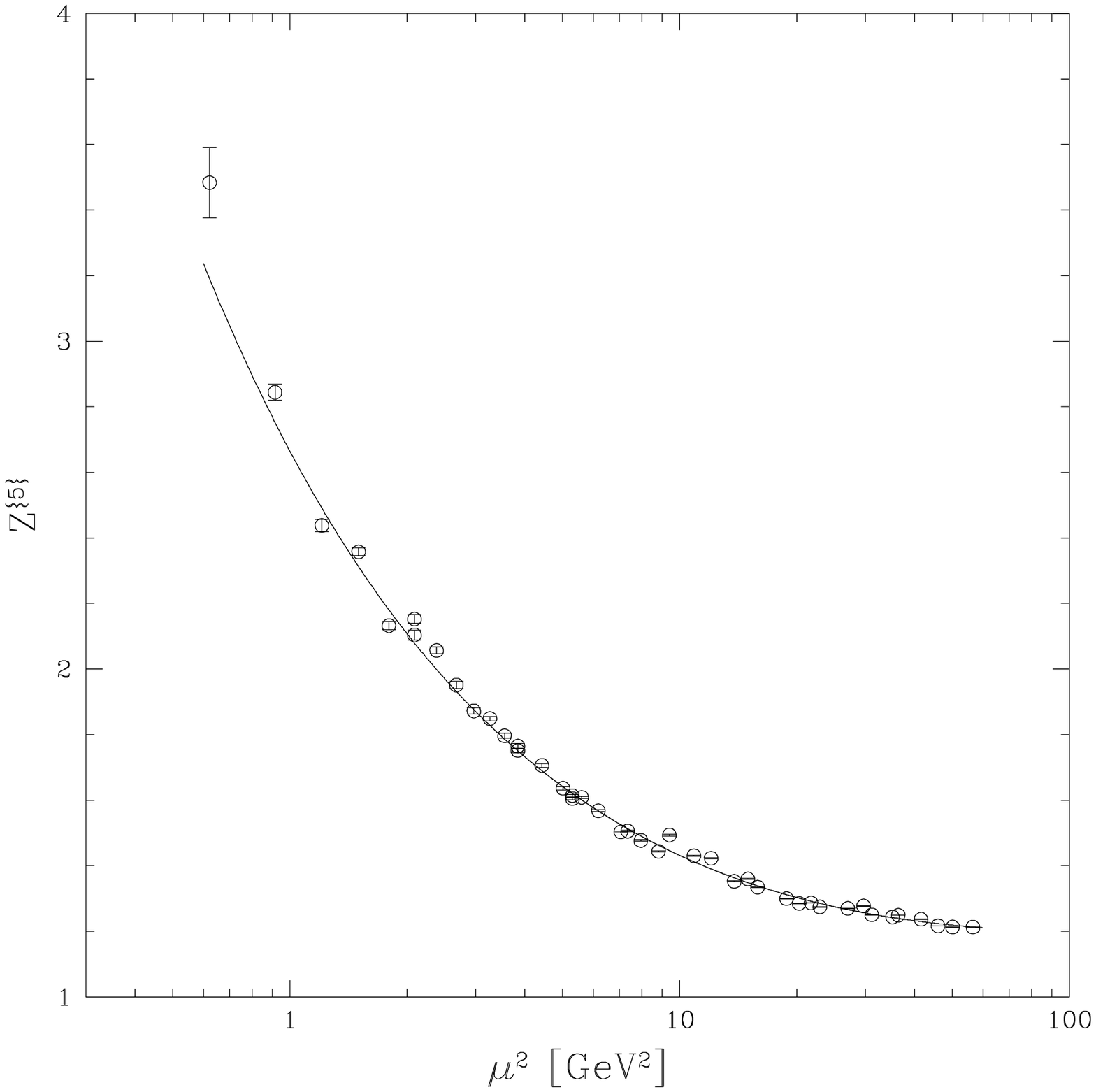,height=10cm,width=10cm}
\vspace*{-0.2cm}
    \caption{The renormalization constant $Z^{\{5\}}$.}
    \label{fig1}
  \end{center}
  \begin{center}
    \epsfig{file=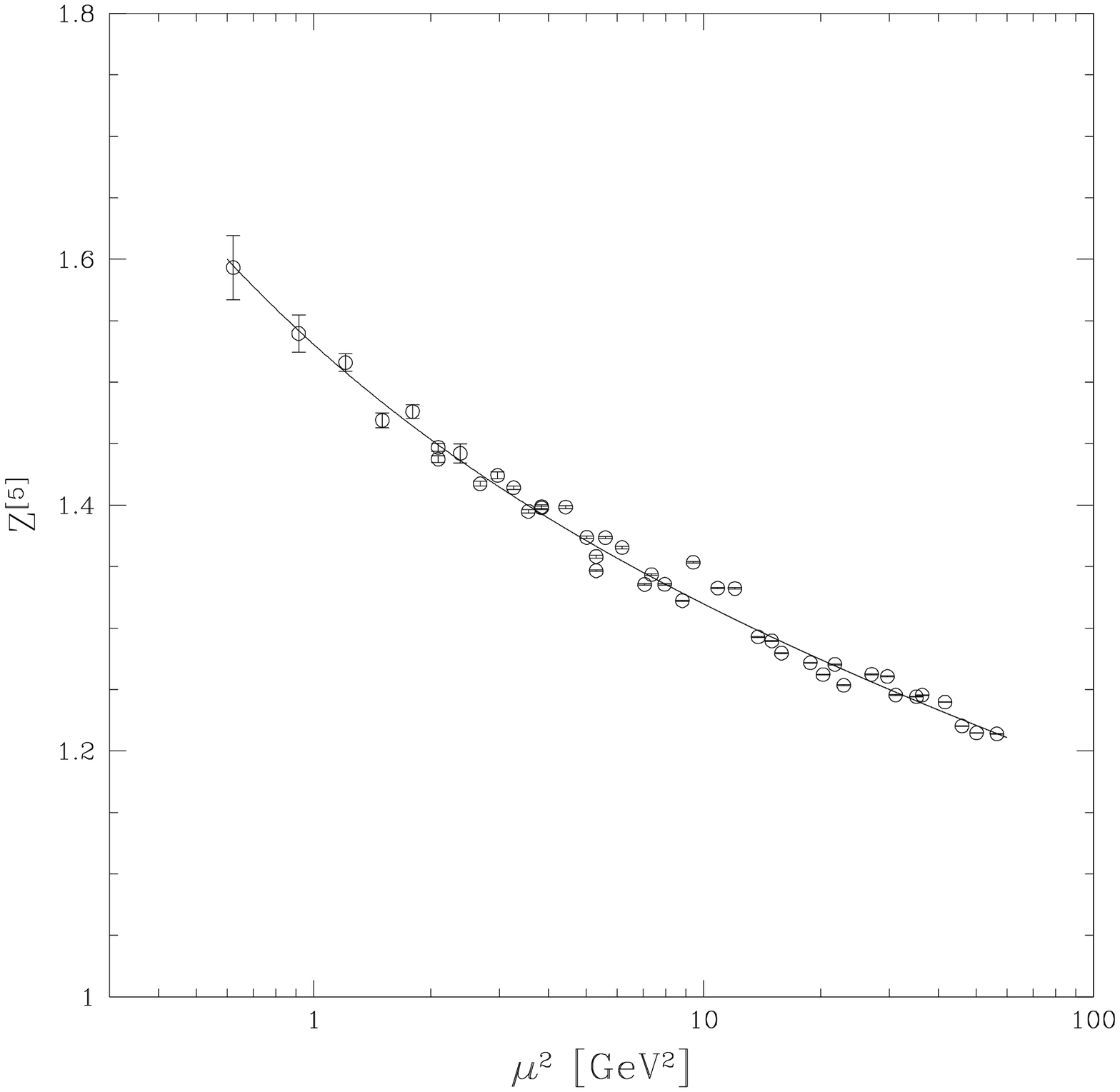,height=10cm,width=10cm}
\vspace*{-0.2cm}
    \caption{The renormalization constant $Z^{[5]}$.}
    \label{fig2}
  \end{center}
\end{figure}

\begin{figure}[htbp]
  \begin{center}
    \epsfig{file=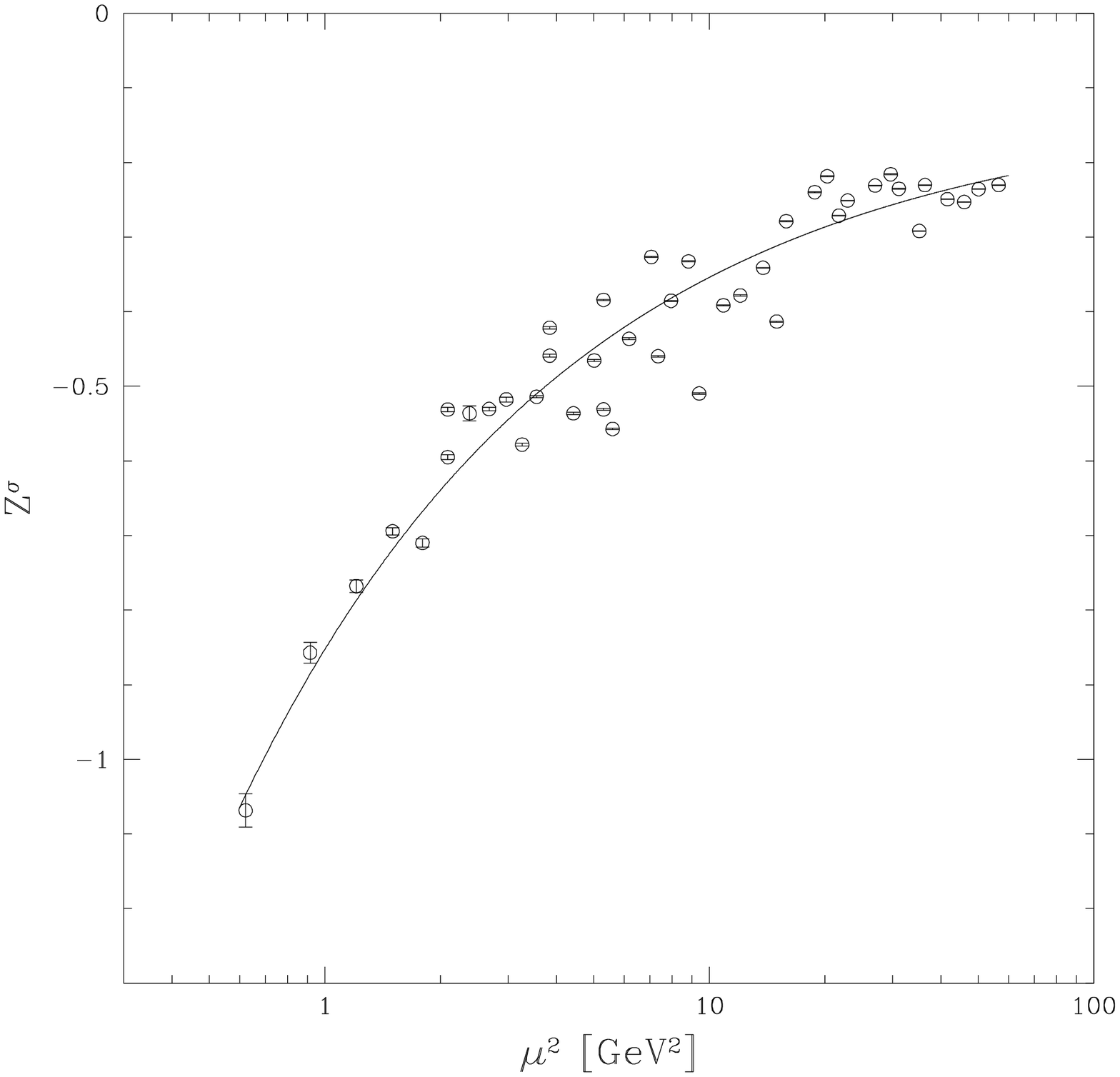,height=10cm,width=10cm}
\vspace*{-0.2cm}
    \caption{The renormalization constant $Z^\sigma$.}
    \label{fig3}
  \end{center}
  \begin{center}
    \epsfig{file=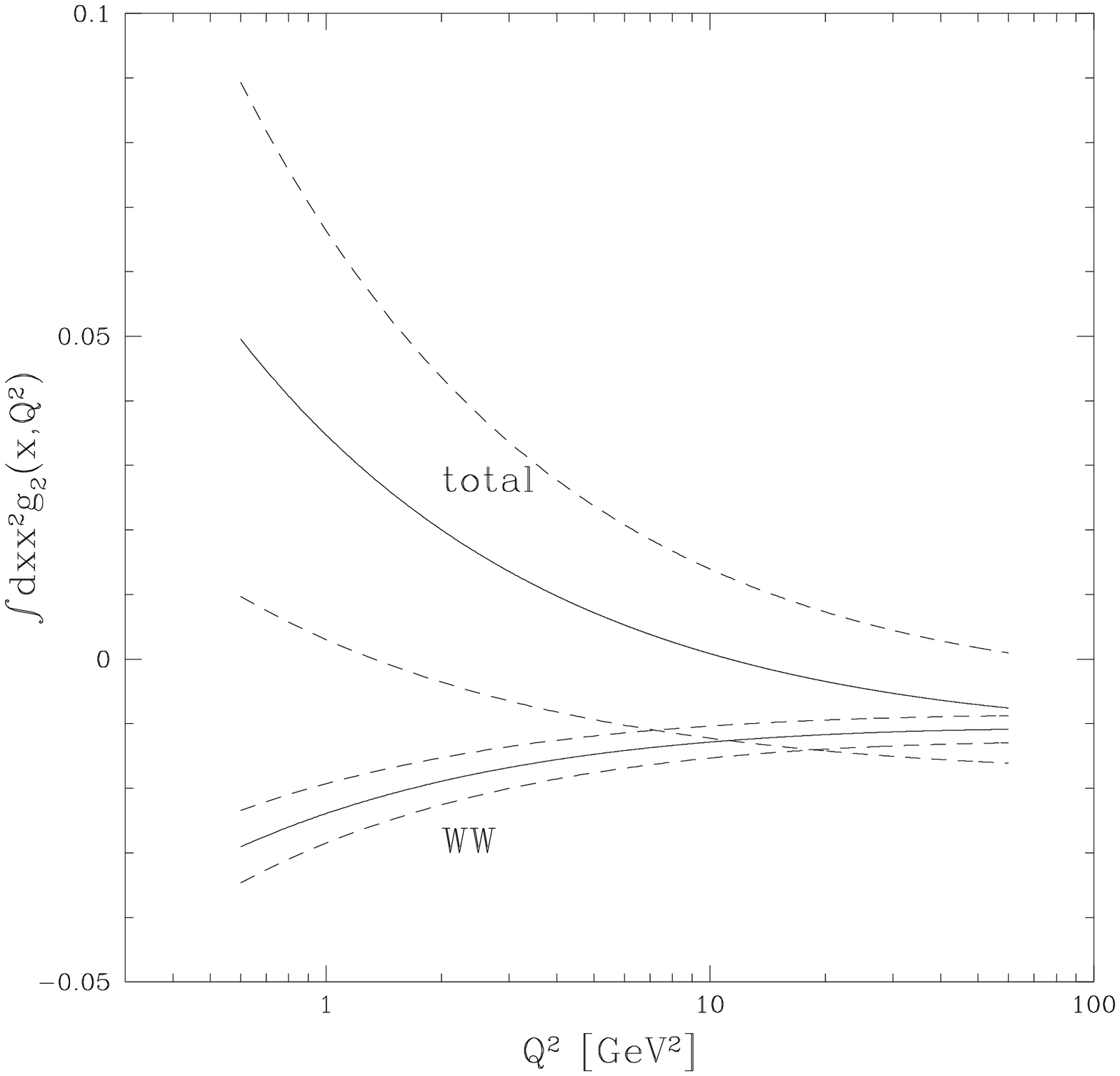,height=10cm,width=10cm}
\vspace*{-0.2cm}
    \caption{The lowest moment of $g_2(x,Q^2)$ for the proton as a function 
of $Q^2$.}
    \label{fig4}
  \end{center}
\end{figure}

The calculation of the renormalization constants follows~\cite{QCDSF3},
supplemented by eqs. (\ref{renorm}) - (\ref{cond2}).
We denote the renormalization
constant of the operator ${\cal O}^{\{5\}}$ by $Z^{\{5\}}$. In
figs.~\ref{fig1} - \ref{fig3} we plot the renormalization
constants as a function of $\mu^2$. The numbers given refer to the MOM 
scheme, and they have been extrapolated to the chiral limit. We fit 
$Z^{\{5\}}$ and  $Z^{[5]}$ by $A + 
B\ln(\mu) + C/\mu$, and $Z^{\sigma}$ by $A + C/\mu$. The result of the fit is 
shown by the solid lines. 
In order to match our results with the perturbatively known Wilson 
coefficients~\cite{Kodaira}, we need to transform the renormalization 
constants to the $\overline{\rm MS}$ scheme. This is done to one-loop 
order~\cite{QCDSF2}.   

\begin{table}[htbp]
\begin{center}
\begin{tabular}{||c|l|l|l|l||} \hline
\raisebox{-10pt}{Operator} & \multicolumn{4}{|c||}{$\kappa$} \\
   & \;0.1515 & \;0.153 & \;0.155 & $\kappa_c=0.1569$ \\ \hline
$ a_2^{(u)} $  & \;0.146(9) & \;0.142(9) & \;0.138(17) & \;\;0.132(23) \\ 
$ a_2^{(d)} $  & -0.032(3) & -0.032(3) & -0.044(9) & \;-0.043(10) \\   \hline
$ d_2^{[5]\,(u)} $  & -0.097(5) & -0.122(6) & -0.177(14) & \;-0.206(18) \\ 
$ d_2^{[5]\,(d)} $  & \;0.018(2) & \;0.021(2) & \;0.032(5) & \;\;0.035(6) \\ 
\hline 
$ d_2^{\sigma\,(u)} $  & -0.74(12) & -0.89(20) & -1.27(110) & \;-1.31(69) \\ 
$ d_2^{\sigma\,(d)} $  & \;0.136(26) & \;0.143(42) & \;0.116(119) & 
\;\;0.137(122) \\ \hline
\end{tabular}
\end{center}
\caption{The unrenormalized, reduced matrix elements $a_2$, $d_2^{[5]}$ and
   $d_2^\sigma$ for $u$ and $d$ quarks separately.}
\label{me}
\end{table}

We are now ready to state our results for $g_2$. At larger values
of $\mu^2$ ($\mu^2
\gap 7 \,\mbox{GeV}^2$) the $\mu$--dependence of $Z^{\{5\}}$ and 
 $Z^{[5]}$ is reasonably well described by the one-loop perturbative result 
(see also~\cite{QCDSF3}), so that we may take $\mu^2 = Q^2$. This choice 
should keep higher-loop corrections to the Wilson coefficients small. In 
fig.~\ref{fig4} we show our new results for the proton. The curve marked 
`total' gives the total contribution of both, twist-two and twist-three 
matrix elements. The curve marked `WW' corresponds to the Wandzura-Wilczek
approximation~\cite{WW}, in which the contribution from twist-three matrix 
elements is discarded. The dashed lines indicate the errors on the curves.
The errors come almost completely from the error of the nucleon matrix 
elements.

Unfortunately, our errors on $d_2$ are yet too large to reach a definite 
conclusion. But we may say already that the effect of mixing is large, 
and of the same magnitude as the twist-three contribution itself. At 
$Q^2 = 5\,\mbox{GeV}^2$ we now find for the proton 
\begin{equation}
\int_0^1 \mbox{d}x x^2 g_2(x,Q^2) = 0.007(15),
\end{equation}
and for the neutron we obtain 
\begin{equation}
\int_0^1 \mbox{d}x x^2 g_2(x,Q^2) = 0.006(7),
\end{equation}
which is not inconsistent with the experimental values~\cite{exp}. In case of
the proton it appears that $d_2$ vanishes at larger values of $Q^2$,
leaving us with the Wandzura-Wilczek result.

We hope to return with more precise numbers in the near future.

\section*{Acknowledgment}

The numerical calculations have been done on the Quadrics computers at
DESY-Zeuthen. We thank the operating staff for support. This work was
supported in part by the Deutsche Forschungsgemeinschaft.

\end{document}